\documentclass[a4paper,fleqn,usenatbib]{mnras}

\usepackage{newtxtext,newtxmath}

\usepackage[T1]{fontenc}
\usepackage{ae,aecompl}

\usepackage{graphicx}	
\usepackage{amsmath}	
\usepackage{amssymb}	

\title[The shadowed ring of DoAr44]{An inner  warp in the DoAr~44 T~Tauri transition disk}

\author[S. Casassus et al.]{
Simon Casassus,$^{1,2}$\thanks{E-mail: simon@das.uchile.cl}
Henning Avenhaus,$^{3,1,2}$
Sebasti\'an P\'erez,$^{1,2}$
V\'{\i}ctor Navarro,$^{1}$ \newauthor
Miguel C\'arcamo,$^{5}$
Sebasti\'an Marino,$^{6}$
Lucas Cieza,$^{2,7}$
Sascha P. Quanz,$^{3}$ \newauthor
Felipe Alarc\'on,$^{1,2}$
Alice Zurlo,$^{1,2,7}$ 
Axel Osses,$^{8}$
Fernando R. Rannou,$^{5}$ \newauthor
Pablo E. Rom\'an,$^{5}$
and Marcelo Barraza$^{1,2}$
\\
$^{1}$Departamento de Astronom\'{\i}a, Universidad de Chile, Casilla 36-D, Santiago, Chile\\
$^{2}$Millennium Nucleus ``Protoplanetary Disks'', Chile\\
$^{3}$ETH Zurich, Institute for Particle Physics and Astrophysics, Wolfgang-Pauli-Str. 27, CH-8093, Zurich, Switzerland\\
$^{4}$Max Planck Institute for Astronomy, K\"onigstuhl 17, 69117 Heidelberg, Germany\\
$^{5}$Departamento de Ingenier\'{\i}a Inform\'atica, Universidad de Santiago de Chile\\
$^{6}$Institute of Astronomy, University of Cambridge, Madingley Road, Cambridge CB3 0HA, UK\\
$^{7}$Facultad de Ingenier\'ia y Ciencias, N\'ucleo de Astronom\'ia, Universidad Diego Portales, Av. Ejercito 441. Santiago, Chile\\
$^{8}$Departamento de Ingenier\'{\i}a Matem\'atica, Facultad de Ciencias F\'{\i}sicas y Matem\'aticas, Universidad de Chile, Beauchef 851, Santiago, Chile
}

\date{Accepted XXX. Received YYY; in original form ZZZ}

\pubyear{2017}

\begin{document}
\label{firstpage}
\pagerange{\pageref{firstpage}--\pageref{lastpage}}
\maketitle

\begin{abstract}
    Optical/IR images of transition disks (TDs) have revealed deep
    intensity decrements in the rings of HAeBes HD~142527 and
    HD~100453, that can be interpreted as shadowing from sharply
    tilted inner disks, such that the outer disks are directly exposed
    to stellar light.  Here we report similar dips in SPHERE+IRDIS
    differential polarized imaging (DPI) of TTauri DoAr~44.  With a
    fairly axially symmetric ring in the sub~mm radio continuum,
    DoAr~44 is likely also a warped system.  We constrain the warp
    geometry by comparing radiative transfer predictions with the DPI
    data in $H$ band ($Q_\phi(H)$) and with a re-processing of
    archival 336~GHz ALMA observations.  The observed DPI shadows have
    coincident radio counterparts, but the intensity drops are much
    deeper in $Q_\phi(H)$ ($\sim$88\%), compared to the shallow drops
    at 336~GHz ($\sim$24\%).  Radiative transfer predictions with an
    inner disk tilt of $\sim 30\pm5$~deg approximately account for the
    observations.  ALMA long-baseline observations should allow the
    observation of the warped gas kinematics inside the cavity of
    DoAr~44.
\end{abstract}




\section{Introduction}

Shadows due to central warps have often been invoked to account for
protoplanetary disk
properties\citep[e.g.][]{Casassus2016PASA...33...13C}, such as the
low-amplitude modulation of the HST images in TW~Hya
\citep{Rosenfeld2012ApJ...757..129R,Debes2017ApJ...835..205D}, or disk
orientation changes in different angular scales \citep[e.g. as in
  GM~Aur, AB~Aur, or MWC~758,][]{Hughes2009ApJ...698..131H,
  Tang2012AA...547A..84T, Isella2010ApJ...725.1735I}. But the link
between shadowing effects and tilted inner disks is best established
in transition disks with large inclination changes, i.e. disks with
radial gaps that separate an inner disk from an outer disk, and where
parts of the outer disk are directly exposed to the star due to a
sufficiently inclined inner disk.  For instance, a tilt of 70~deg may
seem an unlikely phenomenon, yet such is the structure of the
record-sized gap in the protoplanetary disk of Herbig Ae/Be (HAeBe)
HD~142527, as inferred from the identification of the shadows cast by
the inner disk onto the outer ring
\citep{Marino2015ApJ...798L..44M}. With an established disk
orientation, the CO and HCO$^+$ kinematics \citep{Casassus2013Natur,
  Rosenfeld2014ApJ...782...62R} correspond to accretion across the
cavity onto the inner disk, and probably through the warp
\citep{Casassus2015ApJ...811...92C}. This surprising structure is
probably the result of interactions with the low mass companion at
12~AU \citep[mass ratio $\sim$1/10,][]{Biller2012,
  Close2014ApJ...781L..30C, Lacour2016A&A...590A..90L}. Another
dramatic warp, stemming from a sharply inclined inner disk at
$\sim$72~deg, has recently been reported in HAeBe HD~100453
\citep{Casassus2016PASA...33...13C, Benisty2017A&A...597A..42B,
  Long2017ApJ...838...62L,
  Min2017A&A...604L..10M}. One or multiply-sided
  sharp dips are also seen in other sources, such as in GG~Tau~A
  \citep[][]{Itoh2014RAA....14.1438I}, or in HD~135344B
  \citep[][]{Stolker2016A&A...595A.113S}, but their origin is not as
  clearly connected to shadowing from a stable tilted inner disk as in
  the case of the two-sided dips.

The frequency of occurrence of such extreme warps is not yet known,
since the high-contrast requirements are only beginning to be
exploited. Perhaps most disks around small mass ratios binaries ($q
\sim 0.01-0.1$) could warp when the migrating companion crosses the
resonance between its precession period and that of the inner
circumprimary disk \citep{Owen2017MNRAS.469.2834O}. In any case,
thanks to accurate knowledge on disk structure and orientation, the
sharply warped systems represent an opportunity to understand the
physics underlying warped protoplanetary disks in general. What is the
response of circumstellar disks under the forcing of an out-of-plane
companion?  What are the dynamical consequences of shadows deep enough
to cool the gas and reduce the pressure locally?  What can we learn
about disk viscosity?  These long standing questions in disk
hydrodynamics \citep{Papaloizou1995MNRAS.274..987P,
  Nixon_2013MNRAS.434.1946N} require input from concrete observational
evidence.

The general goal of understanding the warp hydrodynamics, and
connections with the origin of transition disk cavities, motivates the
present analysis of a new sharply warped system, this time at T-Tauri
masses.  As part of our survey of Disks ARound TTauri Stars with
SPHERE (DARTTS-S, PI: H. Avenhaus, see Sec.~\ref{sec:obs}), here we
report deep decrements in the stellar infrared radiation that is
reflected, and polarized, on the outer ring of transition disk DoAr~44
(also called ROXs 44, see Fig.~\ref{fig:obs}a). DoAr~44 is located in
the L1688 dark cloud of Ophiuchus \citep{Andrews2011ApJ...732...42A},
so likely close to the distance of $+120.0_{-4.2}^{+4.5}$~pc derived
for the Ophiuchus core by \citet[][]{Loinard2008ApJ...675L..29L}. The
ALMA continuum image \citep{vdm2016A&A...585A..58V} shows a fairly
face-on orientation (inclination of $i =20$~deg), with a smooth ring
and shallow decrements that are strongly modulated by convolution with
the synthetic beam, and which we emphasize here with super-resolution
in non-parametric image synthesis (Fig.~\ref{fig:obs}b,
Sec.~\ref{sec:obs}). With radiative transfer predictions for the
SPHERE+IRDIS polarization images and for the 336~GHz continuum images,
we confirm that the decrements in polarized intensity cannot be
reproduced by radiative transfer effects alone at such low
inclinations. Instead, the data can be interpreted in terms of a
sharply tilted inner disk, in which part of the outer disk is directly
exposed to stellar light (Sec.~\ref{sec:RT}).  We discuss the
observability of the warped kinematics in the cavity of DoAr~44
(Sec.~\ref{sec:discussion}) before summarising our conclusions
(Sec.~\ref{sec:conclusion}).


\section{Observations} \label{sec:obs}

\subsection{Instrumental setups}

\begin{figure*}
\begin{center}
  \includegraphics[width=0.9\textwidth,height=!]{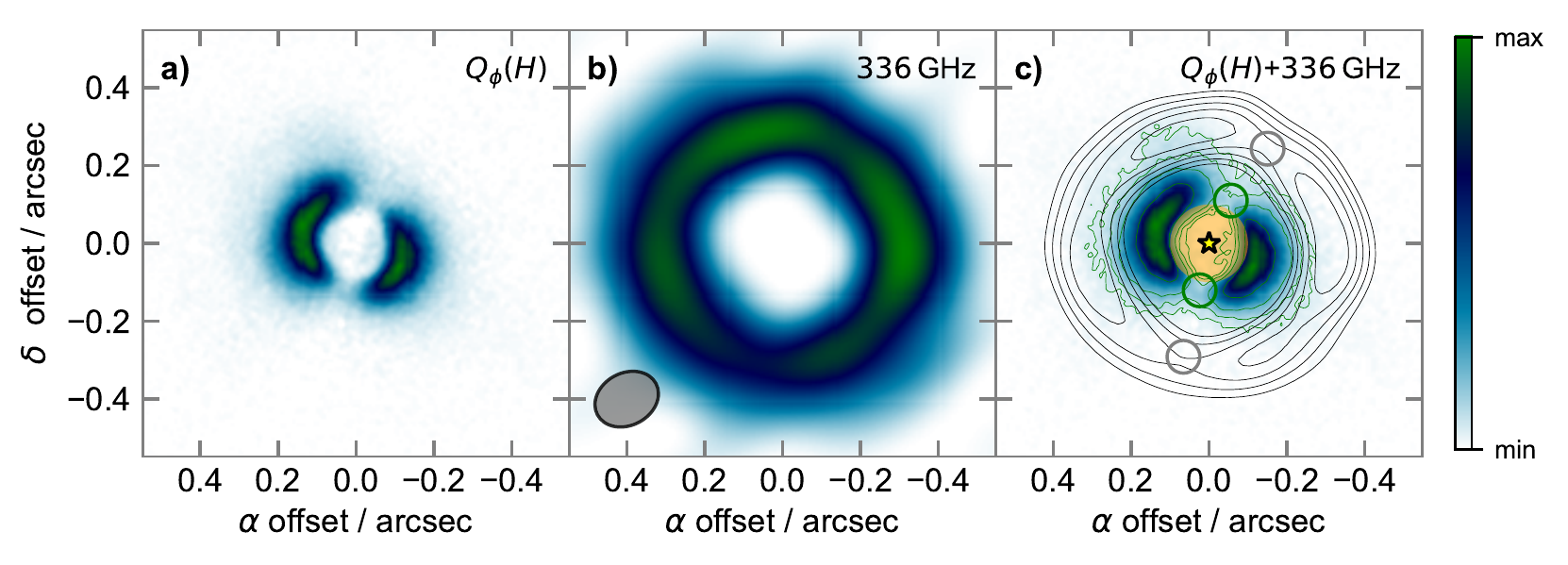}
\end{center}
\caption{ Polarized intensity in $H$-band and deconvolved 336~GHz
  radio continuum from DoAr~44. $x-$ and $y-$ axis show offset from
  the stellar position in the direction of R.A. and Dec., in arcsec.
  {\bf a:} H-band $Q_\phi$ image, with a resolution close to the
  diffraction limit of 49.5~mas.  {\bf b:} 336~GHz continuum image of
  DoAr~44, deconvolved using our {\tt uvmem} algorithm for an
  effective angular shown by the beam ellipse ($0.17\arcsec \times
  0.13\arcsec$, or about $1/3$ the natural-weights clean beam). {\bf
    c:} the 336~GHz image in black contours and overlaid on
  $Q_\phi(H)$. The 336~GHz contours are linearly spaced at fractions
  of [0.5, 0.6, 0.7, 0.8, 0.9] times the peak intensity. We also show
  green contours for $Q_\phi(H)$ at [0.05,0.1,0.2,0.5] times the
  peak. The circular markers indicate the position of the decrements
  along projected circles that best approximate the ring: in green for
  $Q_\phi$, and in grey for the 336~GHz continuum.  The stellar
  position is marked by a yellow symbol.  The semi-transparent orange
  disk indicates the 0.1\arcsec~radius, meant to illustrate the total
  radial extent of the coronagraph.   \label{fig:obs}}
\end{figure*}

DoAr~44 was acquired using SPHERE+IRDIS as part of project {\tt
  096.C-0523(A)}.  The full dataset is described in
\citet[][]{Avenhaus2018arXiv180310882A}. Here we report the DoAr~44
observations acquired in $H-$band, on March 15 2016, with a total
exposure time of $\sim$2560~s. The data were reduced following as in
\citet[][]{Avenhaus2017AJ....154...33A} to produce the $Q_\phi$ linear
combination of the two orthogonal linear polarizations, which
represents an unbiased estimate of the polarized intensity image. This
$Q_\phi(H)$ image is shown in Fig.~\ref{fig:obs}a and
Fig.~\ref{fig:obs}c, where we also compare with a deconvolved image of
the 336~GHz radio continuum.  Details on the radio image synthesis are
provided in Sec.~\ref{sec:IS} and Fig.~\ref{fig:panelL}.



While the continuum ring is fairly smooth and is approximately a
projected circle viewed close to face-on, the $Q_\phi(H)$ ring is
  divided into bipolar arcs, separated by broad and deep intensity
  dips. In this respect, the DoAr~44 decrements are reminiscent of the
  shadows in HD~142527, which are best seen in near-IR imaging, with
  only shallow counterparts in the continuum. On the other hand, their
  coarse and broad shapes in DoAr~44 contrast with the finely drawn
  inner disk silhouettes seen projected on the outer rings of
  HD~142527 and HD~100453. If due to a central warp, this difference
  could reflect the coarser resolution, especially relative to the
  ring radius, or it could represent a difference in warp geometry,
  with a shallower inner disk tilt in DoAr~44.

\subsection{Radio/IR alignment} \label{sec:align}

The usage of additional frames for centering, with four bright spots
imposed by a waffle pattern onto the deformable mirror of SPHERE,
ensures that the SPHERE/IRDIS images are centered on the star to
within a fraction of a detector pixel (so within 12~mas). In turn, the
accuracy of the absolute astrometry of the ALMA data is usually taken
as $\sim$1/10 the synthetic beam, so in this case about 25~mas.

While at the time of writing a parallax for DoAr~44 is not yet
available, the HSOY catalogue \citep[][based on a preliminary GAIA
  release]{Altmann2017A&A...600L...4A} provides astrometric data for
DoAr~44 at epoch 2000: J2000 (16:31:33.4635, -24:27:37.222),
extrapolated from the position at epoch 2015 and accurate to
(4~mas,1~mas), and a proper motion of (-6.329~mas~yr$^{-1}$,
-20.013~mas~yr$^{-1}$), with an error of 2.256~mas~yr$^{-1}$. Since
the date of the ALMA observations is 2014-07-26 \citep[][project ID
  2013.1.00100.S]{vdm2016A&A...585A..58V}, we corrected for proper
motion and shifted the image to epoch 2000 coordinates and centered
the ALMA images on the expected stellar position.



\subsection{Location and contrast of the azimuthal decrements} \label{subsec:shadowloc}

\subsubsection{Azimuthal intensity profiles in polar maps}

\begin{figure}
\begin{center}
  \includegraphics[width=\columnwidth,height=!]{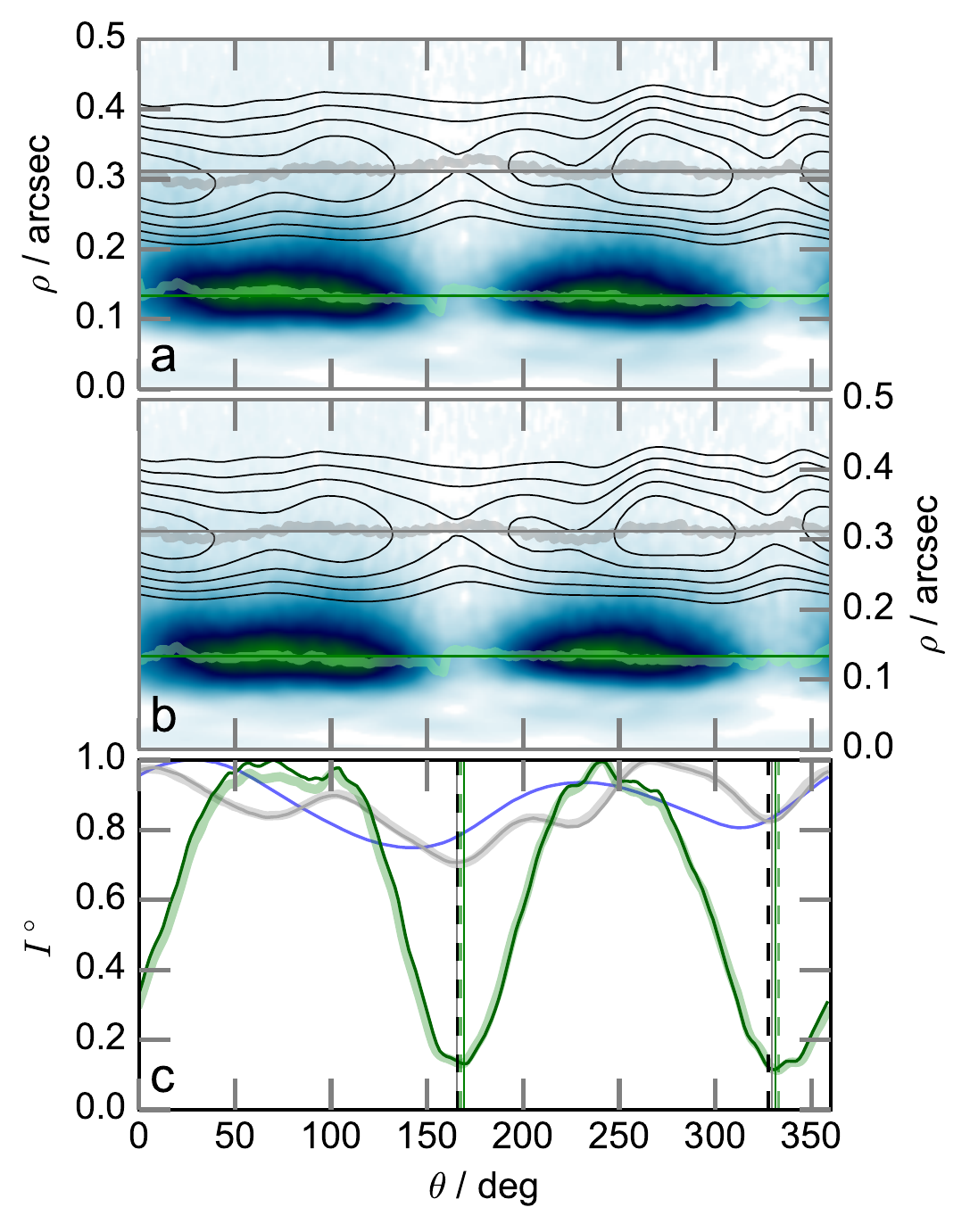}
\end{center}
\caption{Extraction of the ring profiles and impact of the choice of
  polar map origin. {\bf a:} Polar map, centered on the star, of the
  deprojected images. $x-$ and $y-$ axis record azimuth $\theta$, East
  of North, and radial offset.  The thick lines correspond to the
  radial location of the peak intensity, $\rho(\theta)$, in green for
  $\rho_H(\theta)$ in $Q_\phi$, and in grey for $\rho_{336}(\theta)$
  in the 336~GHz continuum. The thin horizonthal lines correspond to
  the averages $\langle \rho_H \rangle = 0.13\arcsec$ and $\langle
  \rho_{336} \rangle = 0.31\arcsec$, with matching colours.  {\bf b:}
  Same as a), but with an origin of the polar coordinates on the ring
  centroids, so offset from the star by $\Delta\rho_c = 8$~mas in the
  direction $\theta_c=197$~deg East of North for the ALMA data, and by
  $\Delta \rho_c =3.8$~mas and $\theta_c = 62$~deg for the DPI data.
  {\bf c:} Resulting ring intensity profiles $I^\circ_H(\theta)$ and
  $I^\circ_{336}(\theta)$, extracted along the constant radii $\langle
  \rho_{H} \rangle$ and $\langle\rho_{336}\rangle$. The thick dashed
  lines correspond to the stellar centroid in a), while the thin solid
  lines correspond to the cavity centroid in b). The location of the
  decrements is marked by the vertical lines, green for $Q_\phi(H)$,
  and grey for 336~GHz, and in dashed lines for the stellar centroid
  (with black instead of grey), and solid for the ring
  centroid.\label{fig:polar}}
\end{figure}

For a quantitative report of the location of the decrements, we first
``deprojected'' the data assuming an inclination $i=20~$deg and a disk
position angle PA=60~deg, East of North, consistent with the molecular
line data presented by \citet{vdm2016A&A...585A..58V}. This merely
corresponds to a stretch of the pixel aspect ratio by a factor
$\cos(i)$ perpendicular to the disk PA.  We then extracted the ring
intensity profile $I^\circ(\theta)$ as a function of azimuth $\theta$
along a circle (Fig.~\ref{fig:polar}c and Fig.~\ref{fig:Iprofiles}a)
that best represents the ring, as we now explain.


\begin{figure}
\begin{center}
  \includegraphics[width=\columnwidth,height=!]{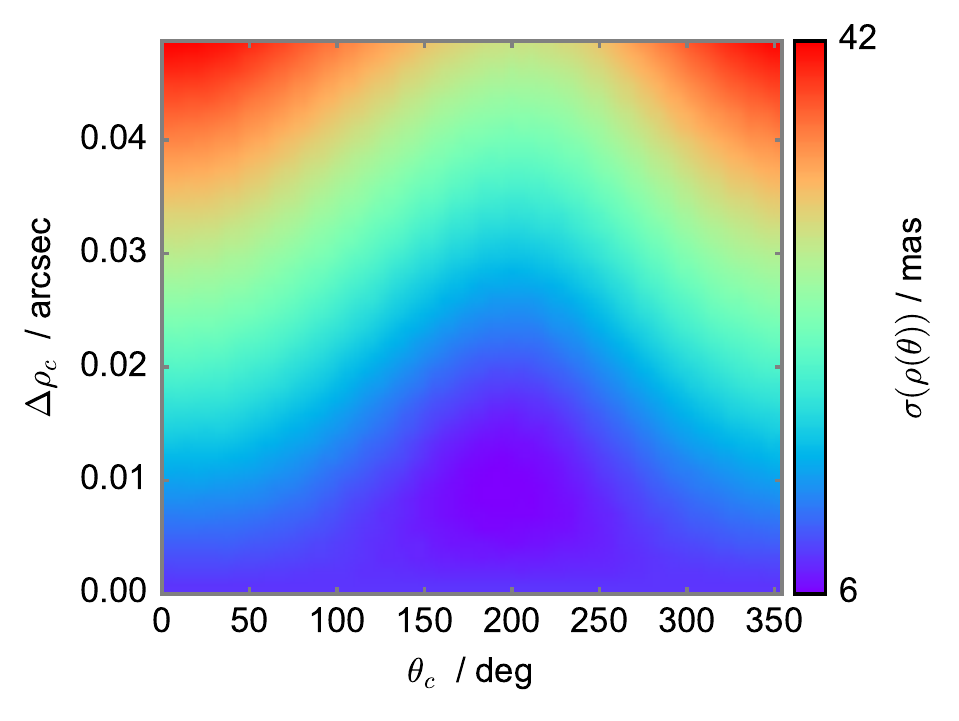}
\end{center}
\caption{Example grid for the location of the 336~GHz ring center. The
  rms dispersion of the ring profile $\sigma(\rho(\theta))$ is shown in
  color stretch, with units of mas, as a function of the direction of
  offset in degrees East of North in $x$-axis, and amplitude of the
  offset in arcsec, $y$-axis.
   \label{fig:offsetgrid}}
\end{figure}

The ring radius profile $\rho(\theta)$ was measured by recording the
radial location of the peak intensity\footnote{since the cavity is not
  a perfect circle, we emphasize the dependency on $\theta$}. The
measurement of $\rho(\theta)$ depends on the choice of origin for the
polar map. Setting the origin to the nominal stellar position
(``stellar centroid'' hereafter) may not be the best choice if the
cavity center is offset from the star, either because of a positional
error, or because of an intrinsic property of the system.  We
therefore optimized the origin so that the shape of the ring is the
closest match to a perfect circle (``ring centroid'' hereafter),
i.e. we minimized the intensity-weighted dispersion in
$\rho(\theta)$. We searched for the cavity centers in each dataset
with a uniform grid in polar coordinates centered on the nominal
stellar position, with a radius of 50~mas. Given a trial origin, we
measured the radial location of the intensity maxima by extracting the
peak along a constant azimuth $\theta$, thus providing profiles for
the ring radius $\rho(\theta)$ and peak intensity $I^p(\theta)$. We
then recorded the root-mean-square dispersion in $\rho(\theta)$, using
$I^p(\theta)$ as a weight function, and produced the map for
$\sigma(\rho(\theta))$ shown in Fig.~\ref{fig:offsetgrid} (for the
case of the radio data). The optimal origin corresponds to the minimum
in $\sigma(\rho(\theta))$.

Provided with an optimal origin, me measure the average ring radius
$\langle \rho(\theta) \rangle$, again using $I^p(\theta)$ as a weight
function. The profiles $I^\circ(\theta)$ were extracted along the
circles $\langle \rho \rangle$, and the position of the decrements
were recorded with the position of the minima in
$I^\circ(\theta)$.

The positions of the decrements are indicated by thin lines in
Fig.~\ref{fig:polar}c (and in Fig.~\ref{fig:Iprofiles}a). The polar
coordinates were then converted back to the sky plane, as indicated
with circular markers in Fig.~\ref{fig:obs}, and as listed in
Table~\ref{table:shadows}. Fig.~\ref{fig:polar} also compares the
polar maps extracted both around the stellar centroid and around the
ring centroids. We see that even such tiny offsets can change the
location of the decrements by up to $\sim$2~deg.

\begin{table}
  \caption{Position and contrast for the intensity decrement in
    $Q_\phi(H)$ and in 336~GHz continuum, 
      resulting from the procedure described in
      Sec.~\ref{subsec:shadowloc}. \label{table:shadows}}
  \begin{center}
  \begin{tabular}{l|ll|ll} \hline
                &  \multicolumn{2}{c}{Dip 1 (at 1~h)}      & \multicolumn{2}{c}{Dip 2 (at 6.5~h)}    \\
                &   $Q_\phi(H)$  & 336~GHz&     $Q_\phi(H)$ & 336~GHz   \\ \hline
  $f_s^a$       &  11$\pm$2\%         &  82$\pm$6\%  &   13$\pm2$\%         & 71$\pm6$\%  \\
    $\theta_s$$^b$  &  331.2    & 329.4  & 169.2         &  165.6 \\
    $\theta^g_s$$^c$  &  336.3  & 328.4  & 169.2         &  165.1 \\
    $r_s^d$     &  0.123        & 0.286  & 0.123         &  0.299 \\ \hline
  \end{tabular}
  \end{center}
  $^a$ minimum intensity over peak  along the  ring \\
  $^b$ PA  of the shadow minimum, in degrees East of North, as viewed on the sky \\
  $^c$ PA  of a  Gaussian centroid for the each shadow, as viewed on the sky, fit within $\pm$5~deg of $\theta_s$ \\
  $^d$ stellocentric separation,  on the sky and in arcsec, of the formal shadow location, defined as the intersection between   direction $\theta_s$ from the ring centroid, with the  best fit circular ring (projected on the sky) 
\end{table}

\subsubsection{Optimal ring centroids and radii}

In the IRDIS images, the $Q_\phi(H)$ ring has a radius $\langle \rho
\rangle_H = 0.131\arcsec$, and its centroid offset by
$\Delta\rho_c$=3.8~mas from the star in the direction
$\theta_c=62$~deg, East of North\footnote{so, if the star is at the
  position given in the HSOY catalogue, or J2000 RA 16:31:33.4635 DEC
  -24:27:37.2215, then the $Q_\phi(H)$ centroid is at J2000 RA
  16:31:33.4637 DEC -24:27:37.2198}. This offset is smaller than the
upper limit uncertainty in the stellar centering (which is itself
better than 12~mas), so that the $Q_\phi(H)$ ring is essentially a
circle centered on the star within the instrumental limitations.

Interestingly, trials on the radiative transfer (RT) predictions
(Sec.~\ref{sec:RT}) result in similar offsets in $Q_\phi(H)$, even
though in this case the cavity is a perfect circle and the stellar
position is known exactly. The RT offset changes with relative
inclination $\xi$ between the inner and outer disks. Examples range
from $(\Delta\rho_c,\theta_c)= (7.5~{\rm mas}, 56.25~{\rm deg})$ at
$\xi=20~{\rm deg}$, to $(3.8~{\rm mas}, 73~{\rm deg})$ at
$\xi=30$~deg, to $(1.3~{\rm mas}, 67.5~{\rm deg})$ at $\xi=40~$deg to
$(0.0~{\rm mas}, 0~{\rm deg})$ at $\xi=60~$deg.

We caution that the $Q_\phi(H)$ gap is very close to the edge of the
coronagraph, such that the shape of the ring could be in part the
result of modulation with the coronograph transfer function, which is
not necessarily centered exactly on the stellar position. Note,
however, that in other disks without central gaps the IRDIS
coronagraph does not result in such deep decrements.

The deconvolved 336~GHz continuum ring has a radius $\langle \rho
\rangle_{336} = 0.309$\arcsec, and its centroid is offset by
$\Delta\rho_c$=8~mas in the direction $\theta_c$=197~deg,
which corresponds to a centroid position of J2000
  RA 16:31:33.4633 DEC -24:27:37.2287. This offset is quite small,
and probably reflects the positional uncertainty of these ALMA data
($\sim$25~mas rms, Sec.~\ref{sec:align}). Trials on the radio RT
predictions resulted in offsets of $\Delta\rho_c=2.5~$mas towards
$\theta_c \sim 140$~deg, and fairly independent of $\xi$.




\subsubsection{Lopsidedness of the 336~GHz ring?}

At 336~GHz the northern side of the disk is somewhat brighter relative
to the south, by $\sim$17\%, as estimated from Fig.~\ref{fig:polar}c
(and Fig.~\ref{fig:Iprofiles}a) by comparing the peak intensity at
$\sim$7~deg with $\sim$200~deg (outside the decrements). Given the
astrometric uncertainty of the radio data (see Sec.~\ref{sec:align}),
this could be due to varying stellar irradiation in an offset ring,
which would require a ring centroid offset towards the South by about
40~mas (if the continuum emission is proportional to the dust
temperature and if $T(r) \propto 1/\sqrt{r}$). The enhanced brightness
of the northern side could also be related to the disk orientation, if
the northern side is the far side. A difference in brightness towards
the exposed inner edge of the ring would require a fairly optically
thick continuum. As discussed in detail in Sec.~\ref{sec:RT}, the RT
predictions (in Fig.~\ref{fig:Iprofiles}) do not support this
interpretation.  Alternatively, the surface density of the ring may
also be moderately lopsided as, for instance, in SR~21
\citep[][]{vandermarel_2015A&A...579A.106V}.


\subsubsection{Depth of the decrements}

The depths of the decrements can be estimated from the contrast ratio
$f_s<1$ between the local minimum and the peak along the ring. While
the IR decrements are quite deep, dropping by $ 100\%-f_s \sim
88\pm1$\%, the radio decrements are rather shallow, with an average
drop of only $\sim 24\pm 6$\% (the uncertainty spans the difference in
the two decrements). Note that on such small scales, with an inner
working angle that borders the coronagraph, the measurement of the
true depth in $Q_\phi$ is affected by the tails of the PSF, so that
the observed values should be considered as lower limits. Likewise,
finer angular resolution in the radio continuum could result in deeper
decrements.

The uncertainties in the depth of each decrement recorded in
Table~\ref{table:shadows} were estimated in the following way. For the
DPI decrements, we used the rms scatter observed in the plateau around
PA$\sim$60~deg, which is at $\sim$2\% of the peak.  The thermal noise
in the radio maps is also very low. The S/N is about $\sim$100 in the
Clean image. Likewise, the S/N is also very high in the deconvolved
model images, about $\sim$50, as estimated by taking statistics on the
RT simulations in 100 different realisations of noise (see
Sec.~\ref{sec:uvbias}). However, this S/N in the radio profiles
ignores the systematics of the deconvolution procedure, so is an upper
limit from the thermal noise only. A lower limit to the S/N can be
estimated from the observed profile itself, if we assume that it
should be the same as the RT predictions, so that any structure in
addition to the shadow decrements is noise. If we take the deepest 
local minimum, at PA=67.7~deg, as a 1~$\sigma$ deviation, then the
noise in the deconvolved radio profile is 6\%. The actual noise level
in the deconvolved radio profiles is probably in-between 2\% and 6\%,
but we record the most conservative value.

%

\subsubsection{Position angle  joining the  decrements, and radio/IR shift $\eta_S$}

\begin{figure}
\begin{center}
  \includegraphics[width=\columnwidth,height=!]{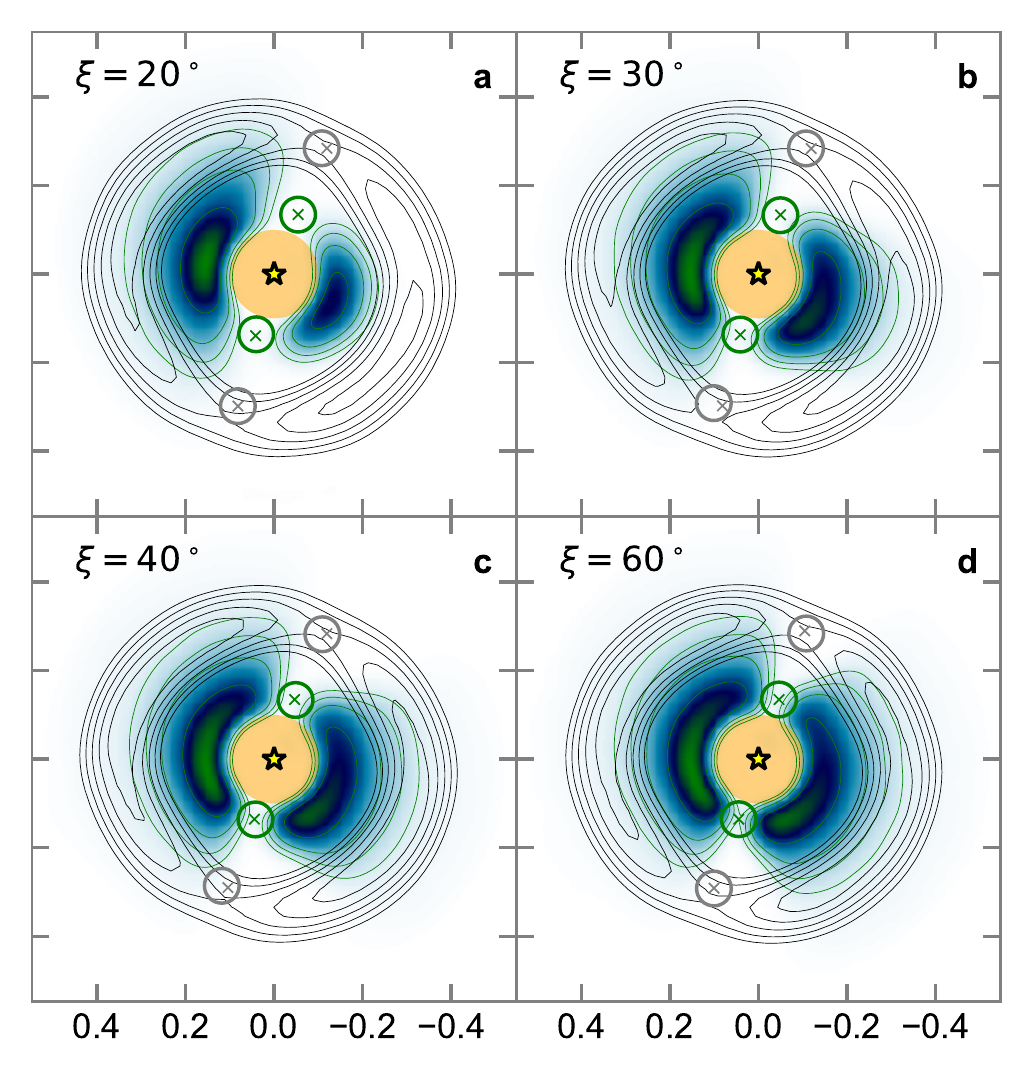}
\end{center}
\caption{ Radiative transfer (RT) predictions for a range of inner
  disk tilt $\xi$, as indicated in each plot. The predictions at
  336~GHz are shown in black contours, with levels as in the
  observations (Fig.~\ref{fig:obs}), and after filtering for the same
  $uv$-plane coverage and image synthesis strategy.  The synthetic
  $Q_{\phi}(H)$, after smoothing to the diffraction limit, and with a
  synthetic coronagraph, is shown in colour stretch, with green
  contours as for the observations. The crosses correspond to the
  position of the decrements when fixing the center of the cavity to
  the stellar position. The semi-transparent orange disk indicates the
  0.1\arcsec~radius of the synthetic coronagraph.
  \label{fig:RTimages}}
\end{figure}

As mentioned above, the position of each decrement, as estimated from
the $I^\circ(\theta)$ profiles, depends on the choice of origin. In
addition, for a tilted inner disk the line joining both shadows is
known to be offset from the star in optically thick scattering on the
disk surface, as in $Q_\phi(H)$, and at finite outer disk inclination
\citep{Marino2015ApJ...798L..44M, Min2017A&A...604L..10M}. We
therefore report the position angle $a$ joining both decrements in
each dataset\footnote{note, however, that the profile still depends on
  the choice of origin. Despite the small offsets, the minima vary in
  position appreciably, as illustrated in Fig.~\ref{fig:RTimages},
  where we also indicate their position with a stellar center. Another
  example of these variations is shown for the observations in
  Fig.~\ref{fig:polar}c.}.  The direction of this line at a given
wavelength can be estimated from the information in
Table~\ref{table:shadows}.

Interestingly, $a$ appears to be very close in the radio and in the
IR.  $a_{336} = -22.0\pm 0.8$~deg in 336~GHz, and $a_{H} = -19.2 \pm
0.5$~deg in $Q_\phi(H)$, as measured East of North, resulting in a
difference of only $\eta_S = a_{336} - a_{H} = -2.8\pm0.9$~deg.  The
uncertainty in $a$ corresponds to the difference with the Gaussian
centroids near the position of each minimum. Note that this
uncertainty does not consider the impact of systematics, such as the
$uv-$plane filtering in the radio data, or varying Strehl ratio in
$Q_\phi(H)$. For instance, the profile for the restored map (which is
comparable to a standard Clean image), shown in Fig.~\ref{fig:polar}c,
corresponds to decrements at a PA of $a = -43.1$~deg, which is clearly
the result of convolution with a coarse and elongated beam.

\subsubsection{Biases from finite angular resolutions and $uv-$plane filtering} \label{sec:uvbias}



In order to estimate the bias from $uv-$plane filtering, we simulated
the same $uv-$coverage on the RT predictions at 336~GHz
(Sec.~\ref{sec:RT}, for a titled inner disk), adding Gaussian noise in
the model visibilities as given in the observations (using the scatter
estimated with CASA task {\tt statwt}).  The input image in native
resolution and example simulations are shown in
Fig.~\ref{fig:panelsimul}.  The result of these Monte Carlo
simulations, with 100 different realisation of noise, is that the
(simulated) shadow PA is $a_{336}=-19.9\pm1.0$~deg, while in the input
RT model, at native resolution, $a_{336}=-15.6$. Decreasing the
relative importance of regularization to $\lambda =0.002$, instead of
$\lambda =0.01$ used in the data (see Sec.~\ref{sec:IS} for
definitions) results in $a_{336}= -16.6\pm1.2$, in closer agreement
with the input. We nonetheless adopt the reconstruction with $\lambda
= 0.01$ for the data since it provides a smoother image while
preserving thermal residuals (see Fig.~\ref{fig:panelL}), at the cost
of biasing the shift in PA between the radio and the IR shadows.

The bias in $\eta_S$ due to convolution with the point-spread-function
(PSF) in $Q_\phi(H)$ is more difficult to estimate because the PSF is
not known as well as in the ALMA data. While flux frames were taken at
the beginning and end of the 2 exposures, the centering of these PSFs
is difficult to ascertain, and they are bound to vary significantly
with Strehl ratio during each exposure. Still, we can estimate the
bias in $a_H$ by comparing the native RT with the diffraction-limit. A
trial on our best synthetic $Q_\phi(H)$, smoothed at the diffraction
limit, results in $a_H=-18.6$~deg. In turn, $a_H$ is preserved in a
synthetic $Q_\phi(H)$ smoothed at 1/10 the diffraction limit, even
though the cavity is smaller than in the coarser case. We conclude
that no significant bias is injected by  smoothing with a circular
PSF at the diffraction limit, but that in practice PSF elongation is
bound to inject some bias in $a_H$.

The observed radio decrements would thus appear to be essentially
coincident with the deeper decrements seen in polarized intensity,
given the small offset in azimuth $\eta_S=-2.8\pm0.9$~deg. While in
the RT predictions at native resolutions we have $\eta_s^m =
+3.0$~deg, the $uv$-plane filtering bias leads to $\eta_s^m = -1.3$,
which is consistent with the observations. We note that any
contribution to $\eta_s$ from the finite cooling timescale of the
shadowed dust, which is responsible for the sub~mm continuum dips,
appears to be negligible.




\section{Radiative transfer model} \label{sec:RT}


\subsection{Parametric model}

It could be thought that the two DPI arcs in DoAr~44 correspond merely
to the double-lobed structure along the disk PA expected in polarized
intensity at finite inclination. An example of such morphologies can
be seen in HD~100546 \citep[][]{Garufi2016A&A...588A...8G,
  Mendigutia2017arXiv171100023M}, at an inclination of 44~deg.  In
DoAr~44, the PA joining the decrements is $a=-19.2$~deg, so quite
close to 90~deg from the disk PA, which is $\sim$60~deg. However, from
Sec.~\ref{subsec:shadowloc} and Table~\ref{table:shadows} we see that
the decrements in $Q_\phi(H)$ do not divide the long axis equally: the
triangle formed by the star and each decrement has an angle at the
stellar vertex of $\omega=160.0\pm0.5$~deg in the plane of the outer
disk. Also, the outer disk has a fairly low inclination of 20~deg (we
confirm this value by checking that the sub-mm ring is indeed a circle
projected by this angle, see Fig.~\ref{fig:polar}). We therefore
constructed an RT model to check if the DPI structure of DoAr~44 could
be understood from simple radiative transfer effects, or else required
a change in disk inclination.

The RT model follows from \citet{vdm2016A&A...585A..58V}. We initially
implemented their parametrization for the dust, including two
populations of grains (small and large), and confirming their SED for
a planar disk. However, by exposing the outer ring directly to the
star, an inner disk tilt boosts the far-IR flux densities by a factor
of 10. Additionally, the two-step-function gas density profile
proposed by \citet{vdm2016A&A...585A..58V} results in a double
concentric ring structure in optical/IR scattered light. We therefore
proceeded to modify this parametric model for a flatter outer ring,
with lower scale height, and with a cubic taper inside the cavity
\citep[so different from the gradual drop implemented
  in][]{vdm2016A&A...585A..58V}. Since the small grains inside the
cavity shield the outer ring, we required very little settling to
raise the temperature of the large grains (i.e. with $\chi = 0.8$, see
Eq.~\ref{eq:scaleheight}).  Another modification is that we used a
Kurucz model atmosphere for the stellar spectrum
\citep[][]{Kurucz1979ApJS...40....1K, Castelli1997A&A...318..841C},
with $T_\mathrm{eff} = 4750$~K, $\log(g)=-4.0$, and with a stellar
radius of $1.5~R_\odot$.  Accretion luminosity was parametrised as in
\citet{vandermarel_2015A&A...579A.106V}. More details on this RT model
are given in Sec.~\ref{sec:detailRT}.

\begin{figure}
\begin{center}
  \includegraphics[width=\columnwidth,height=!]{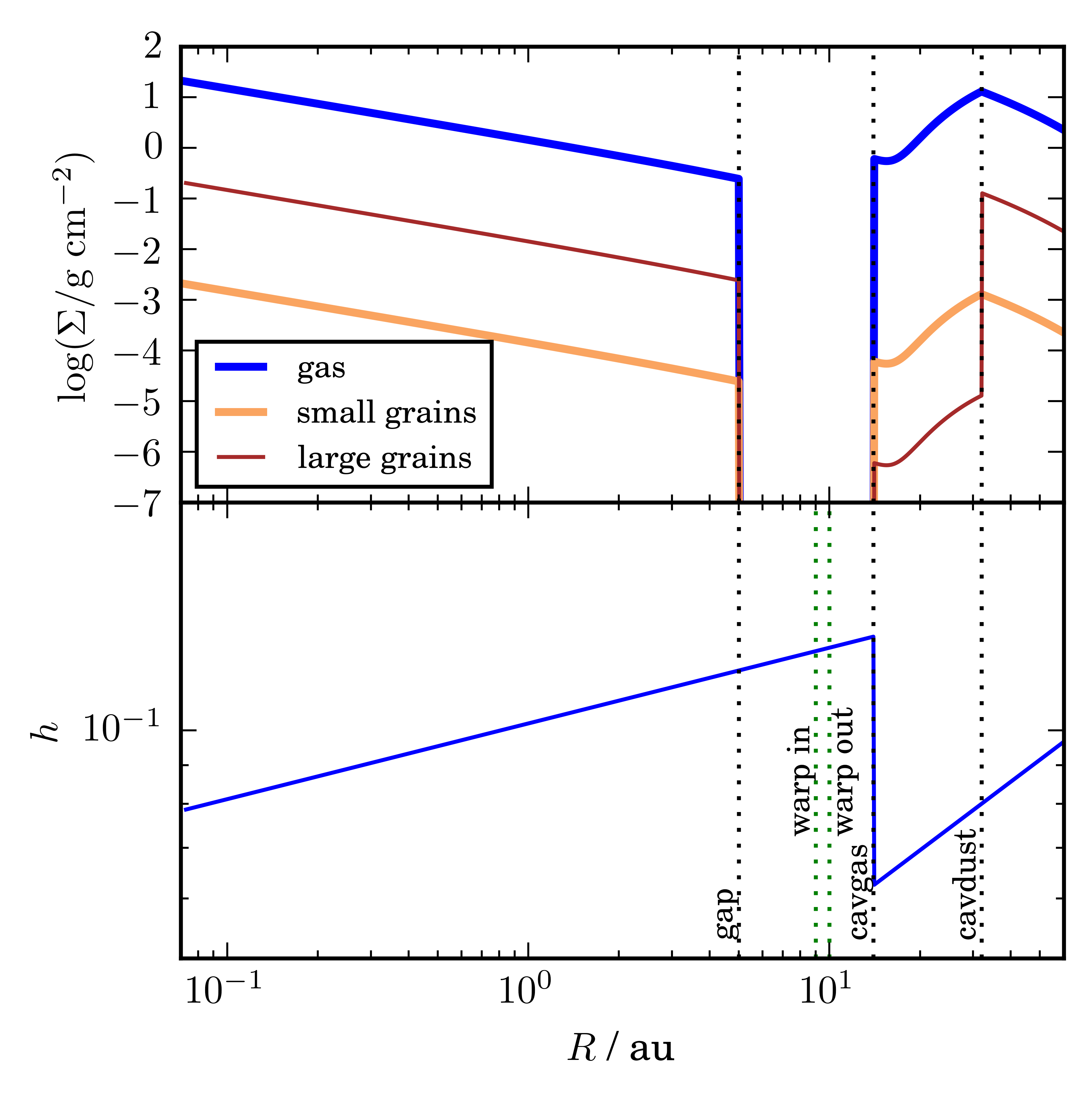}
\end{center}
\caption{ {\bf top:} radial profile for the mass surface density
  $\Sigma(R)$, defined as a function of radius in cylindrical
  coordinates.  {\bf bottom:} scale height $h(R)$, also as a function
  of polar radius. The vertical dashed lines indicate important
  boundary regions (see text for definitions), in
    particular the transition between the inner and outer disk
    inclinations (``warp in, out'') is not constrained by the
    available data.
  \label{fig:profiles}}
\end{figure}

The radial profiles in Fig.~\ref{fig:profiles} summarise the main
features of this parametric model, which we required to be consistent
with the SED (Fig.~\ref{fig:SED} and Table~\ref{table:SED}). The inner
disk is tilted by defining a variable disk orientation that is a
function of cylindrical polar radius, in the frame of the outer disk
\citep[as in][]{Marino2015ApJ...798L..44M,
  Casassus2015ApJ...811...92C}. In our best model (see
Sec.~\ref{sec:alpha}), the relative disk inclination $\xi(R)$ and
orientation $\alpha(R)$ connect linearly from $\xi=0$ at
$R=R_\mathrm{warp\_out}$ to $\xi=30~$deg at
$R=R_\mathrm{warp\_in}$. At the time of writing no high resolution
$^{12}$CO data is available to constrain the location of the
transition in DoAr~44.


The stellar radiation was transferred through the synthetic disk model
using the RADMC3D RT package \citep[][]{RADMC3D0.39}. We calculate
equilibrium grain temperatures and emergent specific intensity maps.
The radio images at 336~GHz were calculated with Henyey-Greenstein
scattering. However, for the calculation of the Stokes images at
1.6$\mu$m we used the scattering matrix (but only incorporating the
small dust as a source of opacity), and post-processed for $Q_\phi$
\citep[as described in][with a focal plane mask of 0.1~arcsec in
  radius]{Avenhaus2017AJ....154...33A}. These predictions are
illustrated in Fig.~\ref{fig:RTimages}.

\begin{figure}
\begin{center}
  \includegraphics[width=\columnwidth,height=!]{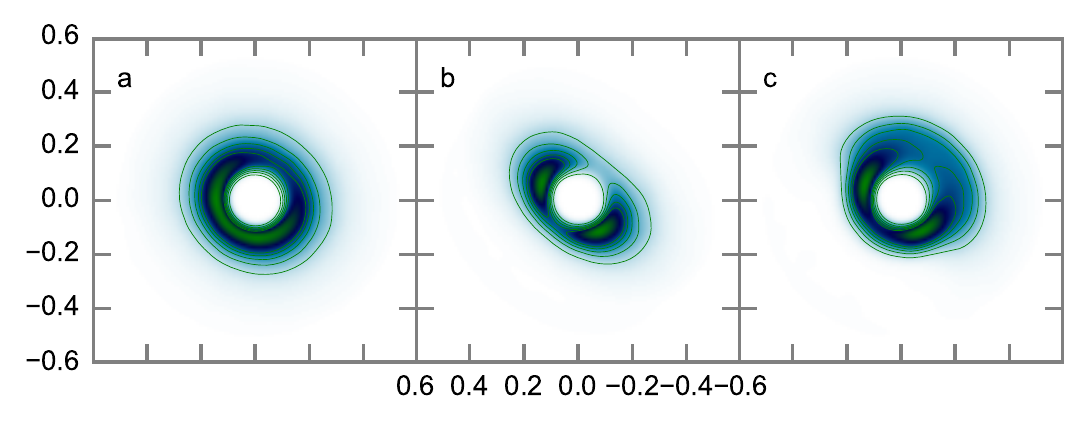}
\end{center}
\caption{ $Q_\phi(H)$ RT predictions without the inner disk, and
  including modulation by a synthetic coronagraph.  {\bf a:} Same as
  our best model (Fig.~\ref{fig:RTimages}b), but without the inner
  disk. We see that the deep decrements are absent. {\bf b:} Same as
  a), but at a disk inclination of 40~deg. The two polarized arcs
  approximate the observations, but the decrements are very
  asymmetric, and they are aligned with the star. {\bf c:} Varying
  inclination, from 20~deg in the outer disk, accounting for the submm
  ring, to 40~deg in the cavity, as required to explain the arcs with
  a high inclination. The morphology is very different from the
  observations.  \label{fig:nowarp}}
\end{figure}

As mentioned above, a modulation of the polarized intensity is
expected at finite disk inclination, resulting in bipolar arcs along
the disk PA. In order to check for this possibility, we calculated
synthetic $Q_\phi(H)$ images without the inner disk, and keeping all
other parameters equal. In Fig.~\ref{fig:nowarp}a, we see that at the
observed inclination of 20~deg, only very shallow decrements are
produced by RT effects alone. Higher inclinations of $\sim$40~deg are
required to reach deep decrements, of $\sim$80\%, as in
Fig.~\ref{fig:nowarp}b. However, these deeper decrements are quite
different from the observations: they are much broader, aligned with
the star, and each have different depths. Most importantly, the
inclination of the submm ring is 20~deg and not 40~deg. In
Fig.~\ref{fig:nowarp}c we consider the possibility of a warp inside
the outer ring, such that the relative disk inclination changes by
60~deg inside the cavity. All of these predictions clearly differ from
the observations, so that we conclude that a tilted inner disk
provides the most simple explanation.

\subsection{Relative disk inclination} \label{sec:alpha}

As explained in this Section, we adopted $\xi=30\pm5~$deg, and an
inner disk PA in the plane of the outer disk of $-90~$deg
(Sec.~\ref{subsubsec:arcs}). Given the outer disk orientation
[$i_2=20~$deg, $\phi_2=60$~deg], this choice corresponds to an inner
disk orientation [$i_1= 29.7$~deg, $\phi_1=134~$deg].  We also
attempted to use directly the location of the decrements, using the
formulae in \citet{{Min2017A&A...604L..10M}}, but the systematics
hamper constraining the inner disk orientation in DoAr~44 with the
later method, as detailed in (Sec.~\ref{subsubsec:geom}).


\subsubsection{Limits from the lengths  of the bipolar DPI arcs} \label{subsubsec:arcs}



\begin{figure}
\begin{center}
  \includegraphics[width=0.9\columnwidth,height=!]{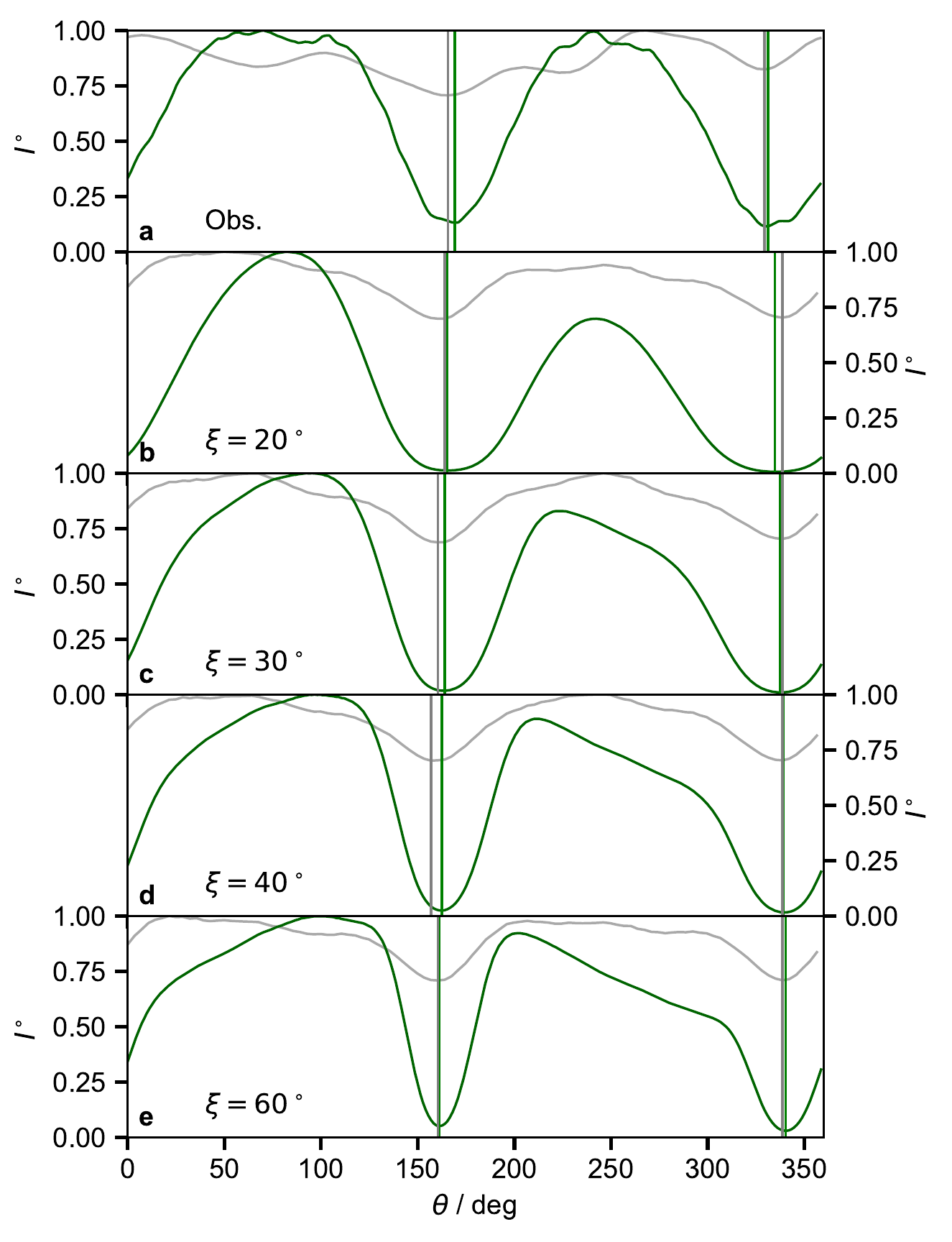}
\end{center}
\caption{Comparison of the observed and ring profiles in intensity as
  a function of azimuth, for different model inclinations. {\bf a:}
  observed profiles (also reproduced in Fig.~\ref{fig:polar}c). The
  solid lines are the intensity profiles extracted from the ring
  centroid (see Sec.~\ref{subsec:shadowloc}), with $I^\circ_H(\theta)$
  for $Q_\phi$ in green, and with $I^\circ_{336}$ for the 336~GHz
  continuum in grey. {\bf b-e:} same as a) but for the RT predictions,
  at inner disk tilt angles (relative inclination between the inner
  and outer disks) of $\xi=20^\circ$, $\xi=30^\circ$, $\xi=40^\circ$,
  and $\xi=60^\circ$. \label{fig:Iprofiles}}
\end{figure}

In Figs.~\ref{fig:RTimages} and \ref{fig:Iprofiles} we note the impact
of the relative disk inclination $\xi$ on the DPI profile of the
ring. Small values of $\xi$ correspond to broader decrements and
asymmetric bipolar emission arcs. We therefore use the length of the
arcs in $Q_\phi(H)$ to constrain the relative disk inclination. We
explored a range of inner disk tilt angles, and found that the lengths
of the arcs in $Q_\phi$ varies significantly with inner disk tilt.
The length of each $Q_\phi$ arcs, East and West, can be estimated with
their standard deviations (after subtracting a linear baseline joining
the two local minima): $s_\mathrm{East} = 38.9$~deg and
$s_\mathrm{West}=32.3$~deg, with a ratio $R_s =
s_\mathrm{East}/s_\mathrm{West} =1.20$.  Large relative inclinations
result in almost equally long arcs, with $R_s = 1.03$ at $\xi=40~$deg,
and $R_s = 0.99$ at $\xi =60~$deg.  For $\xi\lesssim 20~$deg, the arcs
are too short (and the shadows are too wide), with $s_\mathrm{West} =
28.1$~deg. A relative inclination of 30~deg thus seems to be a good
compromise, with $R_s = 1.08$ and $s_\mathrm{West} = 33.5$~deg. We
conclude that the relative disk inclination is likely
$\xi=30\pm5~$deg.

An alternative indicator of model quality could be the ratio of peak
intensities along each arc, but the RT predictions do not take into
account the possibility that the disk is lopsided, which directly
impacts on the azimuthal profiles. The observed peak intensities are
equal on both sides, which would require a very abrupt tilt with $\xi
\gtrsim 60~$deg, and would result in narrow decrements (as in
HD~100453 and HD~142527).

\subsubsection{Geometry inferred assuming perfect circles} \label{subsubsec:geom}

The location of the shadows in $Q_\phi(H)$ can in principle be used to
infer the geometry of the warp. They should approximately correspond
to the intersection of the inner disk midplane with a perfectly
circular outer disk, offset to the altitude of the unit-opacity
surface. The formulae in \citet[][]{Min2017A&A...604L..10M} relate the
inner disk orientation, given by its inclination $i_1$ and its PA on
the sky $\phi_1$, and the outer disk scale height, with the observed
shadow position angle $a$, the angle $\omega$ they subtend relative to
the star (and in the plane of the outer disk), and the stellar offset
$b$ (measured as the intercept of the inner disk PA with the North, at
the stellar right ascension).

In practice, however, the cavity is not perfectly circular, the ring
may be lopsided, and the center of the cavity is difficult to
ascertain. For instance, the small stellar offset from the center of
the cavity measured in $Q_\phi(H)$, of only 3.8~mas
(Sec.~\ref{subsec:shadowloc}), is likely not significant and yet has
an important impact on the inferred inner disk orientation. Another
source of uncertainty is the exact position of the center of the
shadows, especially for such broad decrements as in DoAr~44. The
decrement minimum is not necessarily in the midplane of the inner
disk, since the observed decrements are probably affected by the
coronagraph transfer function, and also by the self cancellation of
extended disk emission in the calculation of $Q_\phi(H)$ at
separations comparable to the PSF
\citep[e.g.][]{Avenhaus2017AJ....154...33A}.

Notwithstanding the above caveats, we solved for the optimal inner
disk orientation in DoAr~44 using the observed values in $Q_\phi(H)$:
$b_\circ = -0.05 \pm 0.012$~arcsec, $\omega_\circ=162\pm 2$~deg, $a =
160.84 \pm0.5$. We sampled the posterior distributions in inner disk
orientation and outer disk scale aspect ratio $h$ using the {\tt emcee}
package \cite[][]{emcee2013PASP..125..306F}. Even such optimistic
uncertainties, which do not include the systematics listed in the
above paragraph, lead to rather loose constraints on the inner disk
orientation: $i_1 =55.6^{+21}_{-11}~$deg, $\phi_1 =
147^{+5.2}_{-17}$~deg, $h = 0.24^{+0.16}_{-0.14}$, where the
uncertainties correspond to the $\pm 1~\sigma$ confidence
intervals\footnote{Note, however, that the posterior distribution in
  $h$ was arbitrarily truncated to $<0.5$}. The distribution of
relative disk inclinations corresponds to $\xi =
56.8^{+12}_{-24}$~deg. Our sign convention for disk inclination is
that a positive value corresponds to a counter-clockwise rotation
along the disk PA \citep[so opposite to the convention used
  in][]{Min2017A&A...604L..10M}. As a comparison point, another
application of the same technique to the parameters in HD~100453, $i_2
= 38$~deg, $\phi_2=142~$deg, $b_\circ= 0.028\pm0.12$, $\omega_\circ=
173 \pm 2$~deg \footnote{the value for $\omega_\circ$ HD~100453 was
  inferred from the optimal solution proposed by
  \citet{Min2017A&A...604L..10M}}, $a_\circ=105.7\pm0.5$~deg
\citep[][]{Benisty2017A&A...597A..42B}, yields $i_1=
-49^{+22}_{-13}~$deg, $\phi_1 = 82^{+15}_{-16}~$deg,
$h=0.176^{+0.14}_{-0.08}$ (or $3.9^{+3.1}_{-1.8}$~au), and a relative
disk inclination $\xi = 74^{+26}_{-17}$~deg. While these values are in
agreement with the inner disk orientation inferred by
\citet{Min2017A&A...604L..10M}, we see that even optimistic errors on
the location of the shadows yield very large uncertainties. In
HD~100453, the star is clearly offset from the cavity center, which
should result in a bias that is not contemplated in this error budget.

%
%
%
%



\section{Discussion: Observing the warped hydrodynamics} \label{sec:discussion}

Accurate knowledge of the warp kinematics is required to guide
research on the physics of warps in protoplanetary disks, and their
possible connection with the origin of large cavities in
general. Following the proposition of \citet{Owen2017MNRAS.469.2834O},
it is possible that some of the large transition disk cavities could
be evacuated by misaligned companions. Indeed,
\citet{Price2018MNRAS.tmp..624P} show, using state-of-the-art 3D
hydrodynamics, that the misaligned companion HD~142527B explains all
key properties of that system, including the inner disk tilt.

Unfortunately the existence of close-in companions is very difficult
to test with high-contrast imaging: even at stellar mass ratios of $q
\approx 0.1$, the detection of HD~142527B was only possible thanks to
instrumental breakthroughs \citep[][]{Biller2012,
  Close2014ApJ...781L..30C}. The ring of DoAr~44 is already at
separations that are close to the inner working angles of even the
latest next-generation AO cameras, hampering further detailed
observations in the optical/IR.

Residual intra-cavity gas is also very difficult to trace in
$^{13}$CO, since there is a significant decrease in gas density inside
the cavity, seen as drop in $^{13}$CO emisison \citep[][their
  Fig.~5]{vdm2016A&A...585A..58V}, whereas $^{12}$CO emission is more
optically thick and thus likely to peak inside the cavity.  We may
thus trace the intra-cavity kinematics in $^{12}$CO.  As reported by
\citet{Salyk2015ApJ...810L..24S}, the inner disk of DoAr~44 is
molecular, with ro-vibrational emission from gaseous CO and H$_2$O
vapor requiring large columns of H$_2$ gas.

The fairly high accretion rate of $dM_\star/dt \sim
10^{-8}$~M$_{\odot}$~yr$^{-1}$ in DoAr~44
\citep[][]{Manara2014A&A...568A..18M} would deplete the total gas mass
of the inner disk, of $4~10^{-6}~$M$_\odot$, in less than 500~yr
(assuming a standard gas to dust ratio of 100 and the model of
Sec.~\ref{sec:RT}). This depletion time is about a couple of orbits of
the outer ring, which would make the observation of the inner disk a
very unlikely phenomenon.  Thus, material must cross the cavity and
replenish the inner ring, much like in HD~142527. In steady state, the
radial and infalling velocity component is directly linked to the
surface density profile, $\Sigma(r) = (dM_\star/dt) / (2 \pi r v_r)$,
so that the gap should correspond to fast radial infall.

\begin{figure}
\begin{center}
  \includegraphics[width=\columnwidth,height=!]{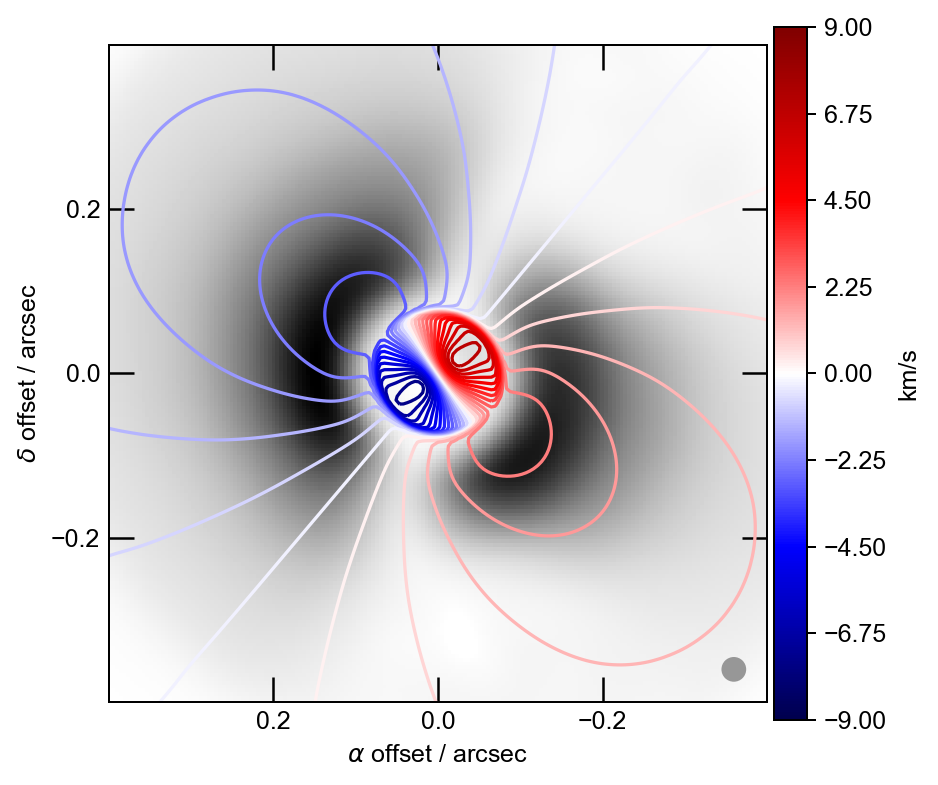}
\end{center}
\caption{ Expected CO(6-5) velocity centroid overlaid on the $Q_\phi$
  prediction (after convolution with a 30~mas beam), for an inner-disk
  tilt of 30~deg. The contour colours correspond to the velocities
  given by the wedge, in km~s$^{-1}$.  \label{fig:CO65}}
\end{figure}

Even without the incorporation of a radial velocity component, i.e. in
pure Keplerian rotation, the proposed inner disk tilt in DoAr~44
should be detectable with ALMA in $^{12}$CO(6--5), as shown in
Fig.~\ref{fig:CO65} \citep[we confirmed that the integrated line
  profile of the synthetic $^{13}$CO(3-2) is consistent with that
  reported by][]{vdm2016A&A...585A..58V}. The higher J lines are
preferred to minimize absorption by diffuse ISM screens, especially in
the direction of Ophiuchus at $v_\mathrm{LSR} \sim 3-4~$km~s$^{-1}$,
so close to the systemic velocity of DoAr~44 ($v_\mathrm{LSR} \sim
4~$km~s$^{-1}$).  The RT predictions for CO(6-5) have been smoothed to
a 30~mas beam, which is the best possible with ALMA in band~9.  We see
that the tilt of the inner disk should correspond to a shift in PA,
and to a discontinuity in the velocity pattern.  Any residual
absorption by the diffuse screen near systemic velocity would not
affect the higher velocity channels. Incidentally, the apparent PA of
the inner disk is not coincident with the direction of the optical/IR
shadows, an effect already noted by \citet{Min2017A&A...604L..10M}.

The intra-cavity velocity field of DoAr~44 is unlikely to be
Keplerian, and should bear similarities to HD~142527. We have already
mentioned the need for infall to replenish the inner disk. In
addition, in a continuous warp material must be accelerated from one
plane to the other. So inside a warp there should be a velocity
component orthogonal to the plane of the disk $v_{\rm warp}$.
Interestingly, the inclusion of $v_{\rm warp}$ improves the model of
HD~142527, the only warp for which we have data, with fairly
point-symmetric high velocity ridges connecting the cavity `twist'
with the inner disk, which are missing in a model without $v_{\rm
  warp}$ \citep[the `slow warp model' in][see their Figs~1~b and
  1~d.]{Casassus2015ApJ...811...92C}. We expect similar features in
DoAr~44.  Resolved observations in CO(6-5) would thus inform on the
detailed structure of the warp: the range of stellocentric radii where
the disk breaks from one plane to the other, and the radial velocity
field. We may even expect azimuthal modulations in HCO$+$ \citep[as in
  the HCO$^{+}$(4-3) filamentary structures seen in
  HD~142527][]{Casassus2013Natur}, if the gas connects the two
orientations along streamers.


\section{Conclusions} \label{sec:conclusion}


New DPI imaging of T Tauri star DoAr~44 with SPHERE+IRDIS reveals deep
azimuthal decrements in $Q_\phi(H)$. These dips have a counterpart in
the radio continuum at 336~GHz. The observed DPI decrements are much
deeper than in the radio: while the intensity drops by $\sim$88\% in
$Q_\phi(H)$, the radio dips in the deconvolved images are relatively
shallow, with a drop of $\sim$24\% at 336~GHz. The location of the
optical and radio decrements are coincident, within fairly narrow
uncertainties, and including image synthesis biases. A parametric
model with a central warp provides a simple explanation for these
features. We conclude that an inner disk tilt of $30\pm5$~deg accounts
for the observations.


\section*{Acknowledgments}

We thank the referee for constructive comments and improvements to the
article, and also Drs. Martin Altmann, Ren\'e Mendez and Edgardo Costa
for pointing us to the HSOY catalogue. This work makes use of archival
ALMA data from project 2013.1.00100.S.  Financial support was provided
by Millennium Nucleus RC130007 (Chilean Ministry of Economy), and
additionally by FONDECYT grants 1171624 and 1171841.  This work used
the Brelka cluster, financed by FONDEQUIP project EQM140101, and
hosted at DAS/U. de Chile.  SP acknowledges CONICYT-Gemini grant
32130007. H.A. acknowledges support from NCCR PlanetS supported by the
Swiss National Science Foundation. PR acknowledges CONICYT PAI project
79160119.









\appendix

\section{ALMA image synthesis} \label{sec:IS}

In their original resolution the ALMA 336~GHz continuum observations
show interesting structure along the ring
\citep[][]{vdm2016A&A...585A..58V}. However, the beam is comparable to
the ring radius, and strongly smooths out the intensity profile
along the ring. Since the source is relatively bright, with a peak
specific intensity of $\sim$0.29~Jy~beam$^{-1}$ in natural weights
(with a beam of $0.35\arcsec\times 0.24\arcsec$) and a dynamic range
$\gtrsim 125$, we attempted to super-resolve the continuum data using
our {\tt uvmem} package \citep[][]{Casassus2006,
  Casassus2015ApJ...811...92C, Carcamo2018A&C....22...16C}, which is
part of the family of algorithms based on maximum-entropy
regularization. Here we used the GPU adaptation from
\citet{Carcamo2018A&C....22...16C}, and regularized by minimizing the
Laplacian of the model image, with the following objective function:
\begin{equation}
  L = \frac{1}{2} \sum_{k=0}^{N} \omega_k \left\| V^\circ_k - V^m_k\right\|^2 +  \lambda \sum_{ij} (\Delta p_{ij})^2. \label{eq:L} 
\end{equation}
In Eq.~\ref{eq:L}, $N$ is the total number of observed visibilities
$V^\circ$, each with a weight $\omega_k = 1/\sigma_k^2$. $V^m_k$ are
the model visibilities, calculated on the model image $I^m(x_i,y_j)$,
which is itself directly related to the free parameters $p_{ij} =
I^m(x_i,y_j)/\sigma_D$, where $\sigma_D$ is the thermal noise in the
natural-weights dirty map. The dimensionless parameter $\lambda$
controls the relative importance of the regularization term, which we
chose here as the image Laplacian:
\begin{eqnarray}
  \Delta p_{ij} &= &  p_{i-1,j}+p_{i+1,j}+p_{i,j-1}+p_{i,j+1}-4p_{i,j},  \\
   & \approx & \left( \left. \frac{\partial^2 p(x,y)}{\partial x^2}\right|_{x_{ij}} + \left. \frac{\partial^2 p(x,y)}{\partial y^2}\right|_{x_{ij}} \right) \delta x\delta y,
\end{eqnarray}
where $\delta x$ and $\delta y$ are the image pixel sizes. 

\begin{figure*}
\begin{center}
  \includegraphics[width=0.9\textwidth,height=!]{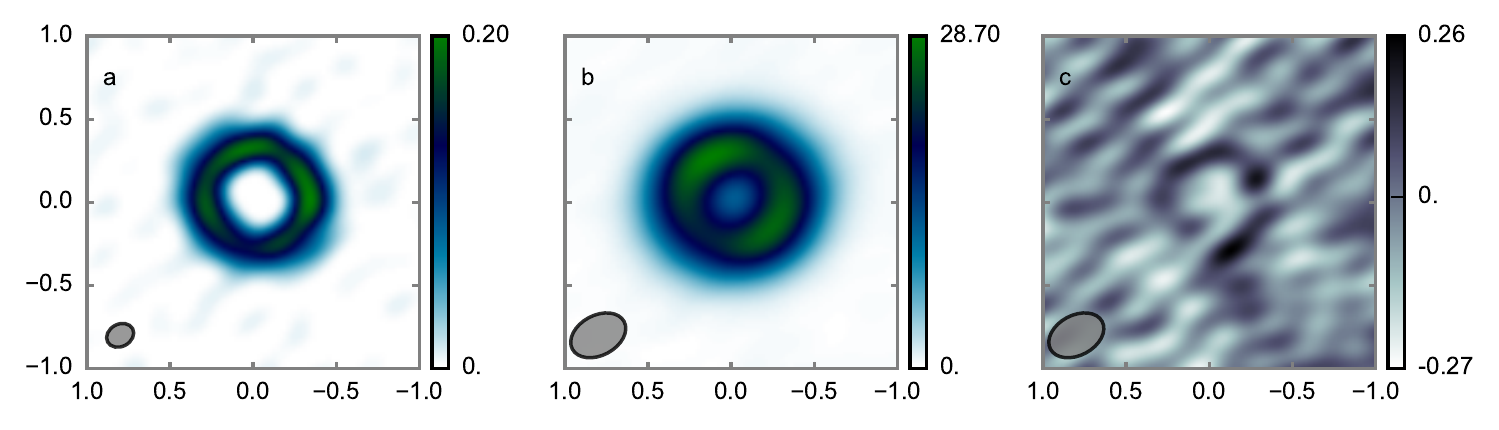}
\end{center}
\caption{Summary of the non-parametric image synthesis. {\bf a:} Model
  image, obtained by minimizing the least squares differences with the
  observed visibility data, and regularized by minimizing the total
  image Laplacian. Units are in mJy~pix$^{-1}$, for a pixel size of
  0.02\arcsec.  The effective angular resolution is given by the beam
  ellipse $(0.17\arcsec \times 0.13\arcsec)$ {\bf b:} Restored image,
  obtained by convolving the model image with the clean beam (in
  natural weights), and by adding the residuals shown in {\bf
    c)}. Units for b) and c) are in mJy~beam$^{-1}$, with a beam of
  $0.35\arcsec\times 0.24\arcsec$. \label{fig:panelL}}
\end{figure*}

After self-calibration of the full continuum dataset \citep[ALMA
  project ID 2013.1.00100.S,][]{vdm2016A&A...585A..58V}, we obtained
the image shown in Fig.~\ref{fig:panelL}. We chose $\lambda = 0.01$,
which is the highest value that preserved thermal residuals. The
super-resolved version of the 336~GHz continuum data provides improved
visualization of the structure modulating the ring, at levels that are
closer to its intrinsic contrast.

\begin{figure}
\begin{center}
  \includegraphics[width=\columnwidth,height=!]{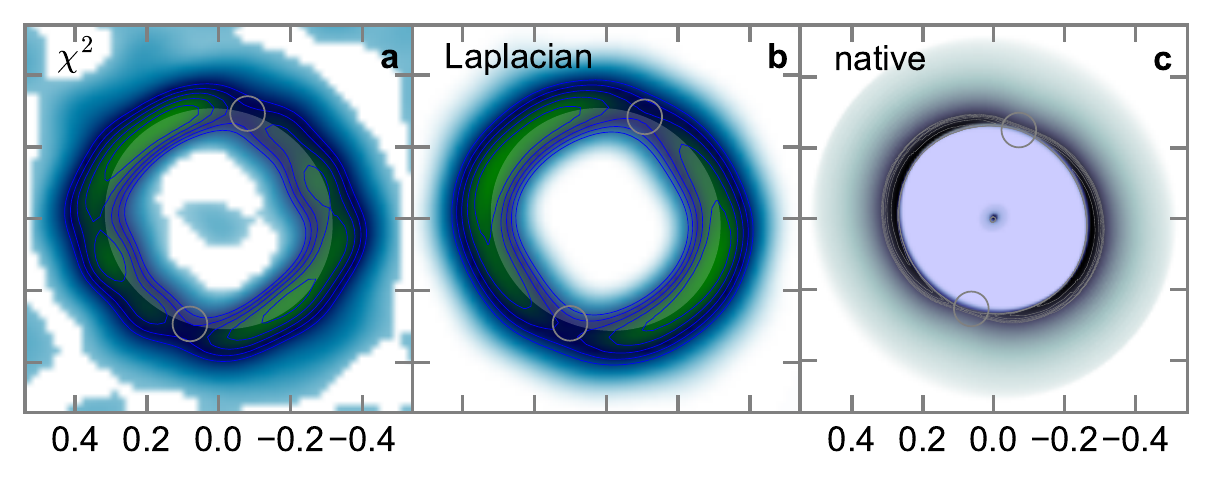}
\end{center}
\caption{Example simulations of image deconvolution on synthetic
  visibility data. The semi-transparent ellipses correspond to the
  optimal projected disk that fits the outer rings.  {\bf a:} Model
  image obtained by minimizing $\chi^2$, without regularization except
  for image positivity. {\bf b:} Model image obtained with Laplacian
  regularization, as Fig.~\ref{fig:obs}b. {\bf c:} Radiative
  transfer prediction input to the simulations in a) and
  b). \label{fig:panelsimul}}
\end{figure}

As explained in Sec.~\ref{subsec:shadowloc}, the impact on the
position of the radio decrements due to the $uv$-plane filtering and
our choice of regularization was estimated with Monte Carlo
simulations on the RT model. Fig.~\ref{fig:panelsimul} gives an
example, for one realization of noise, for both plain $\chi^2$ (with
positivity), and with Laplacian regularization.

Similarly, we also estimate the effective angular resolution of the
model image by simulating on a spike, whose flux is set to the peak
intensity along the outer ring in Jy~beam$^{-1}$, including the
addition of noise. After fitting an elliptical Gaussian, we obtained
an effective angular resolution of $(0.17\arcsec \times
0.13\arcsec)$. This angular resolution depends on the level of the
signal, but in general it varies between $1/3$ and $1/2$ times the
clean beam in natural weights \citep[][]{Casassus2015ApJ...811...92C,
  Carcamo2018A&C....22...16C}.

\section{Parametric RT model } \label{sec:detailRT}

\begin{figure}
\begin{center}
  \includegraphics[width=\columnwidth,height=!]{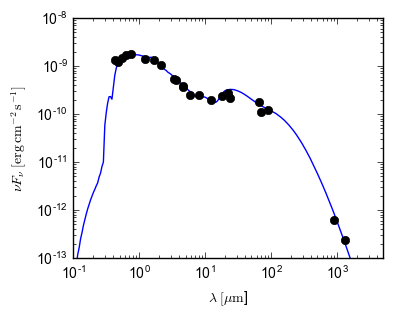}
\end{center}
\caption{ Observed and model (solid line) spectral energy distribution
  for DoAr~44.  The photometry from Table~\ref{table:SED} was
  dereddened by $A_V = 2.2$mag.
\label{fig:SED}}
\end{figure}

\begin{table} \small
\caption{Observed total flux densities in DoAr~44 \label{table:SED}.}
\centering
\begin{tabular}{c|c|c}
                & $F_\nu$ (mJy)                           & $\lambda$ ($\mu$m)                    \\  \hline
USNO$^a$        &   11.9, 34.63, 16.9,  &    0.43 , 0.54 , 0.47, \\ 
                &   61.04, 119.35       &     0.62 , 0.75 \\
2MASS$^b$       &  323, 515, 602        &     1.23, 1.66, 2.16              \\ 
IRAC$^c$        & 620, 540, 489, 673             & 3.6 , 4.5 , 5.8 , 8             \\ 
WISE$^c$        & 601.5, 582.9,       & 3.4 , 4.6 ,              \\ 
          & 778.4, 2055      &  12 , 22             \\ 
MIPS$^c$        & 1690, 2560                     & 24 , 70                         \\ 
AKARI$^{d,e}$       & 1459, 3874, 3638               & 18 , 65 , 90                    \\ 
ALMA$^f$        & 183.77                               & 880                                   \\ 
JCMT$^g$        & 105                                  & 1300                                 
\end{tabular}

$^a$ \citet{2013AJ....145...44Z}
$^b$ \citet{Vizier2014yCat.2332....0E}  
$^c$ \citet{2012yCat.2311....0C}  
$^{d,e}$ \citet{2010A&A...514A...1I, Vizier2010yCat.2298....0Y}
$^{f}$ \citet{Andrews2007ApJ...671.1800A}
$^{g}$ \citet{vdm2016A&A...585A..58V} 
\end{table}

Fig.~\ref{fig:SED} and Table~\ref{table:SED} summarise the available
information from the spectral energy distribution
(SED). Fig.~\ref{fig:SED} also includes the RT prediction from the
warped model that we propose for DoAr~44. Here we proceed to described
the model in detail. 

The gas density profile is parametrised in cylindrical coordinates,
\begin{equation}
n_\mathrm{H2}(R^{\prime\prime},\theta^{\prime\prime},z^{\prime\prime})
= \frac{\Sigma(R^{\prime\prime})}{\sqrt{2\pi} r^{\prime\prime} h
  (R^{\prime\prime})} \exp\left[ -\frac{1}{2}\left(
  \frac{z^{\prime\prime}}{R^{\prime\prime}
    h(R^{\prime\prime})}\right)^2\right],
\end{equation}
where the double primes denote coordinates in a transformed coordinate
system $\mathcal{S^{\prime\prime}}$.  This frame is obtained by
rotating the outer disk frame $\mathcal{S}$, whose $(\hat{x},\hat{y})$
plane correspond to the midplane. We rotate $\mathcal{S}$ by $\phi(r)$
around $\hat{z}$, to an intermediate frame $\mathcal{S^\prime}$, and
then by $\xi(r)$ around $\hat{x}^\prime$. The disk scale height is
\begin{equation}
  h(R^{\prime\prime}) = \chi   h_\circ  \, (R^{\prime\prime}/R_{h_\circ})^\beta, \label{eq:scaleheight}
\end{equation}
where $\chi$ is a settling parameter \citep[as
  in][]{Andrews2011ApJ...732...42A}. The surface density profile is
the usual,
\begin{equation}
  \Sigma(R^{\prime\prime}) = \omega_\mathrm{taper} \Sigma_\circ  \, (R^{\prime\prime}/R_{\Sigma_\circ})^{-1} \exp\left(-\frac{R^{\prime\prime}}{R_{\Sigma_\circ}}\right),
\end{equation}
in which 
$\omega_\mathrm{taper}(R^{\prime\prime})$ implements a gradual gas
drop inside the cavity, 
\begin{equation}
\omega_\mathrm{taper}(R^{\prime\prime}) =   
\delta_\mathrm{cavity} +  (1 - \delta_\mathrm{cavity})  \left( \frac{ R^{\prime\prime} - R_\mathrm{cavgas}}{R_\mathrm{cavdust} - R_\mathrm{cavgas}}\right)^3,
\end{equation}
with $\delta_\mathrm{cavity} = 10^{-2}$ inside $R_\mathrm{cavgas} < R
< R_\mathrm{cavdust}$, and $\delta_\mathrm{cavity} = 1$
elsewhere.

We considered two dust populations, with sizes corresponding to small
and large grains. The small dust is assumed to be
  coupled with the gas, while the larger grains are affected by
moderate settling, corresponding to $\chi \lesssim 1$. Large grains
are assumed to be rarefied inside the cavity, by a factor
$\delta_\mathrm{dust}$. Grain optical properties were calculated using
70\% astro-silicates, and 30\% amorphous carbon, with optical
constants from \citet{Draine2003ApJ...598.1026D} and
\citet{Li_Greenberg_1997A&A...323..566L}, and mixed using the
Bruggeman formula for a solid density of 2.9~g~cm$^{-3}$.  The
resulting grain opacity is 3.1~g~cm$^{-2}$ at 1300~$\mu$m.

Parameter values are given in Table~\ref{table:RTparams}. We caution
that this parametric model is meant to demonstrate that the data can
be interpreted in terms of a tilted inner disk, with constraints on
the inner disk orientation based on the morphology of the shadows
(Sec.~\ref{subsec:shadowloc}). Hence we do not explore parameter space to
place uncertainties on each parameter.



\begin{table}
  \small
  \caption{\label{table:RTparams} Set of parameters for the radiative transfer model. }
  \begin{center}
    \begin{tabular}{lrrrrr} \hline
                                          & $h_\circ$  &   $\beta$ &  $\delta_g$ &  $\delta_\mathrm{cav}$ & $R_{h_\circ}$ \\ \hline 
    $R_\mathrm{sub}  \le R < R_\mathrm{gap}$   &  0.12 & 0.1 & $10^{-3}$ &  1 &  $R_\mathrm{gap}$  \\
    $R_\mathrm{gap} < R <  R_\mathrm{cavgas}$  & 0.12  & 0.1 &  $10^{-10}$   & 1      &   $R_\mathrm{gap}$\\
    $R_\mathrm{cavgas} < R <  R_\mathrm{cavdust}$ &  0.08  &  0.3 & 1  &  $10^{-2}$ & $R_\mathrm{cavdust}$  \\
    $R_\mathrm{cavdust} < R < R_\mathrm{out}$     &  0.08   & 0.3  &  1  &  1  &  $R_\mathrm{cavdust}$\\ \hline
  \end{tabular}

  \smallskip

    \begin{tabular}{cccc} \hline
          $R_\mathrm{sub}^a$ &  $R_\mathrm{gap}^a$ &  $R_\mathrm{warp in}^{a,b}$ &   $R_\mathrm{warp out}^{a,b}$   \\ \hline
       0.07   & 5  & 9 & 10  \\\hline
  \end{tabular}

    \begin{tabular}{cccc} \hline
       $R_\mathrm{cavgas}^a$ & $R_\mathrm{cavdust}^a$ & $R_\mathrm{out}^a$ & $R_{\Sigma_\circ}^a$  \\ \hline
        14 & 32 & 60 & 25 \\\hline
  \end{tabular}

  \smallskip
  \begin{tabular}{lcccc} \hline
        &  $(a_\mathrm{min},a_\mathrm{max})^c$   & $\chi$   & $f^d$  & $\delta_\mathrm{dust}^e$ \\   \hline
    small dust  &  (0.005,1)    & 1   &  0.01  & 1          \\
    large dust  & (0.005,1000)  & 0.8 &  0.99  & $ 10^{-4}$  \\ \hline
  \end{tabular}\\
  $^a$ radius units are in AUs\\
  $^b$  this narrow warp is chosen  where the gas density is set $\sim$0, out of lack of observational data  \\
  $^c$ range of grain radii, given in $\mu$m, with a distribution $a^{-3.5}$  \\
  $^d$ mass fraction of each dust population\\
  $^e$ dust depletion factor for $R < R_\mathrm{cavdust}$ \\
  $^f$ the gas surface density ar $R_{\Sigma_\circ}$ is $\Sigma_\circ = 60$~g~cm$^{-2}$\\
  \end{center}
\end{table}

%
%
%

%


\bsp	
\label{lastpage}
\end{document}